\documentclass{appolb}
\usepackage{graphicx}
\usepackage{multirow}
\begin{document}
\title{Study of some (non-)conventional mesons in the framework of effective models.%
\thanks{Presented at Excited QCD 2020, 2-8 February 2020, Krynica-Zdr\'oj, Poland.}%
}
\author{M. Piotrowska
\address{\textit{Institute of Physics, Jan Kochanowski University,}\\ \textit{ul. Uniwersytecka 7, 25-406, Kielce, Poland. }}
\\
}
\maketitle
\begin{abstract}
The main aim of our study is to understand the nature of some conventional and non-conventional mesonic states by applying effective QFT models. We start from the relativistic Lagrangians containing a unique $q\bar{q}$ seed state which is strongly coupled to the low-masses decay products of the original state. We find out that some states may appear as a dynamically generated companion poles of the heavier $q\bar{q}$ mesons. In particular we show that $K^*_0(700)$ is a companion pole of the well-known $K^*_0(1430)$ resonance, $X(3872)$ emerges as a (virtual) companion pole of $\chi_{c1}(2P)$, and the puzzling $Y(4008)$ is not a real state, but a spurious enhancement which appears when studying the state $\psi(4040)$.
\end{abstract}
\PACS{PACS numbers come here}
  
\section{Introduction}
Mesons listed in the PDG are mostly conventional quark-antiquark objects \cite{pdg}. Yet, other non-conventional mesonic states such as tetraquarks, molecules, hybrids and glueballs are also possible \cite{Brambilla}. Intense research, both on theoretical and experimental levels, could not yet give unambiguous explanation on the nature of some of these states, even if many progresses have been made. 

In these proceedings based on  \cite{kappaMilena, psi4040Milena, XMilena} we present a short review on the status of some non-conventional mesons belonging to scalar and vector sectors. First, we discuss how to construct our theoretical models. Then we present the main results for the $K^*_0(1430)$, $\psi(4040)$ and $\chi_{c1}(2P)$ systems, where the effects of dynamical generation of poles are clearly visible.
 
The mechanism of generation of `additional companion poles' is rather simple and was applied in numerous works in the field, e.g in \cite{a0, Coito, Boglioneprd}. Let us first consider a (bare) seed state which corresponds (in the non-interacting limit) to a well-established quark-antiquark meson. This single $q\bar{q}$ seed state is included in the Lagrangian and couples strongly to some lower in mass ordinary mesons (as for instance pions and kaons). As a result of strong interaction quantum fluctuations emerge. The propagator of the original state is dressed by the mesonic quantum loops. In consequence, the pole of the seed state is shifted in the complex plane and some changes in the shape of the original spectral function are observed. Moreover, in case of  strong coupling of the standard $q\bar{q}$ state to its decay products, one might observe an additional companion pole. This means that at the end, out of one seed, two poles appear. In some cases this additional pole might be assigned to a new resonance. This phenomenon can explain the nature of some non-conventional mesons. 

Along this line we show that $K^*_0(700)$ emerges as a companion pole of the conventional $K^*_0(1430)$ resonance and $X(3872)$  can be understood as a (virtual) companion pole of the $c\bar{c}$ state $\chi_{c1}(2P)$. Moreover, we show that enigmatic state $Y(4008)$ is only an enhancement which emerges when studying $\psi(4040)$ resonance, and should not be regarded as a real state.

\section{Theoretical formalism}
In our studies we use an effective relativistic Lagrangians which describe the decays of a single $q\bar{q}$ seed state into lighter $q\bar{q}$ mesonic pairs. Depending on the considered system, the Lagrangians take different forms which are listed in Table \ref{tablag}. 
\begin{table}[h]
\renewcommand{\arraystretch}{1.53}
\par
\makebox[\textwidth][c] { 
\par%
\begin{tabular} 
[c]{c|c|c|c}\hline \hline
Seed&$J^{PC}$&$q\bar{q}$ state &Lagrangian\\ \hline
$K^*_0$&$0^{++}$&$K^*_0(1430)$& $\mathcal{L}_{K^*_0}=aK^{*-}_0 \pi^0 K^++bK^{*-}_0 \partial_{\mu} \pi^0 \partial^{\mu} K^++\ldots$\\ \hline
$\psi$&$1^{--}$&$\psi(4040)$&$\mathcal{L}_{\psi}=ig_{\psi} \psi_{\mu}(\partial^{\mu}D^+D^--\partial^{\mu}D^-D^+)+ \ldots$\\ \hline
$\chi_{c1}$&$1^{++}$&$\chi_{c1}(2P)$&$\mathcal{L}_{\chi_{c1}}=g_{\chi_{c1}}\chi_{c1, \mu}(D^{*0,\mu}\bar{D}^0+D^{*+, \mu}D^-)+\ldots$\\ \hline \hline
\end{tabular}
}\caption{\label{tablag} Lagrangians for the resonances $K^*_0(1430)$, $\psi(4040)$ and $\chi_{c1}(2P)$.}%
\end{table}
The dots in the presented expressions stand for further combinations of the isospin multiplets. Notice that the Lagrangian corresponding to $K^*_0(1430)$ state is somewhat specific since it consists of two terms, one with derivative and one without it. 

As a next step we present the theoretical formulas of the decay widths that are obtained by using an ordinary Feynman rules. The corresponding expressions for each system together with the main decay channels and examples of the Feynamn diagrams are shown in Table \ref{tabwidth}.
\begin{table}[h]
\renewcommand{\arraystretch}{1.63}
\par
\makebox[\textwidth][c] { 
\par%
\begin{tabular}
[c]{c|c|c|c}\hline \hline
State & Decay& Decay width $\Gamma(m)$ & Example of\\ 
 & channel & (theoretical formula)& Feynman Diagram\\ \hline
\raisebox{5.5ex}{$K^*_0(1430)$}&\raisebox{5.5ex}{$\pi K$}&\raisebox{5.5ex}{$3\frac{|\vec{k}|}{8 \pi m^2}[a-b\frac{m^2-m^2_{\pi}-m_{K}^2}{2}]F_{\Lambda}$}&\includegraphics[width=0.215 \textwidth]{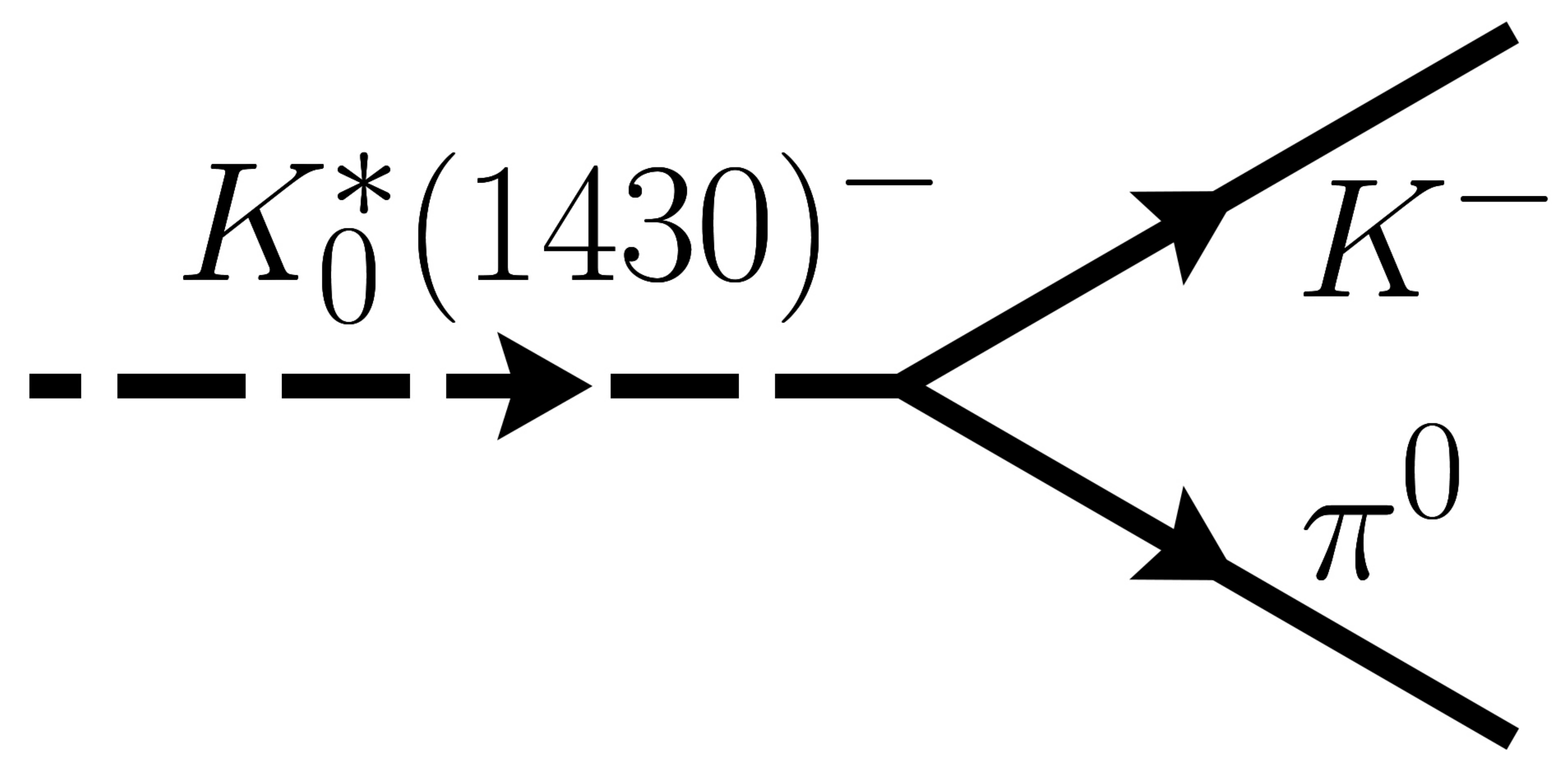}\\\hline
\raisebox{5.5ex}{}&\raisebox{5.5ex}{$DD, D_sD_s$}&\raisebox{5.5ex}{$\frac{|\vec{k}|^3}{6 \pi m^2_{\psi}}g^2_{\psi DD} F_{\Lambda}$}&\includegraphics[width=0.215 \textwidth]{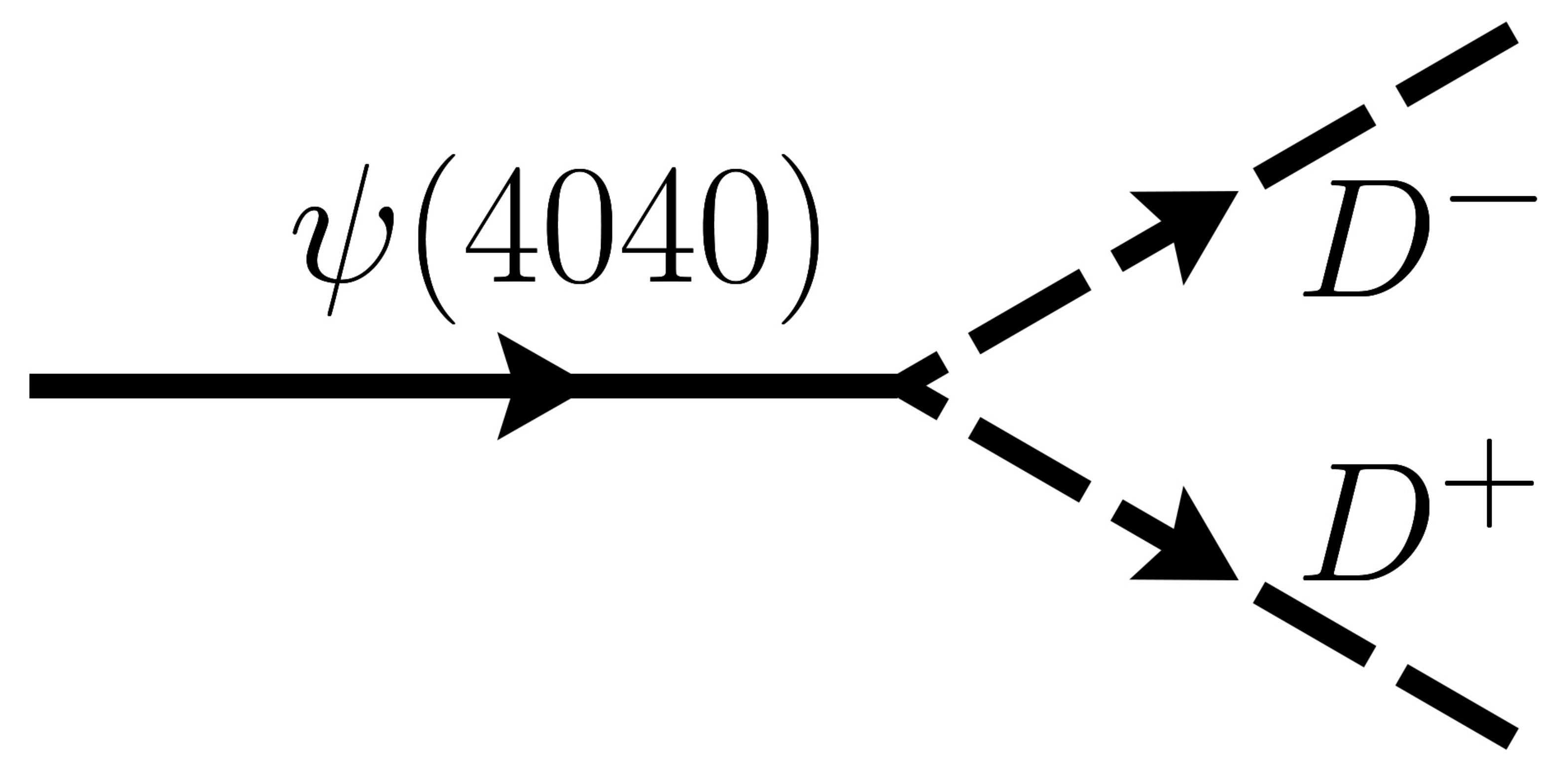}\\
\raisebox{5.5ex}{$\psi(4040)$}&\raisebox{5.5ex}{$D^*D, D_s^*D_s$}&\raisebox{5.5ex}{$\frac{2}{3}\frac{|\vec{k}|^3}{\pi}g^2_{\psi D^*D}F_{\Lambda}$}&\includegraphics[width=0.215 \textwidth]{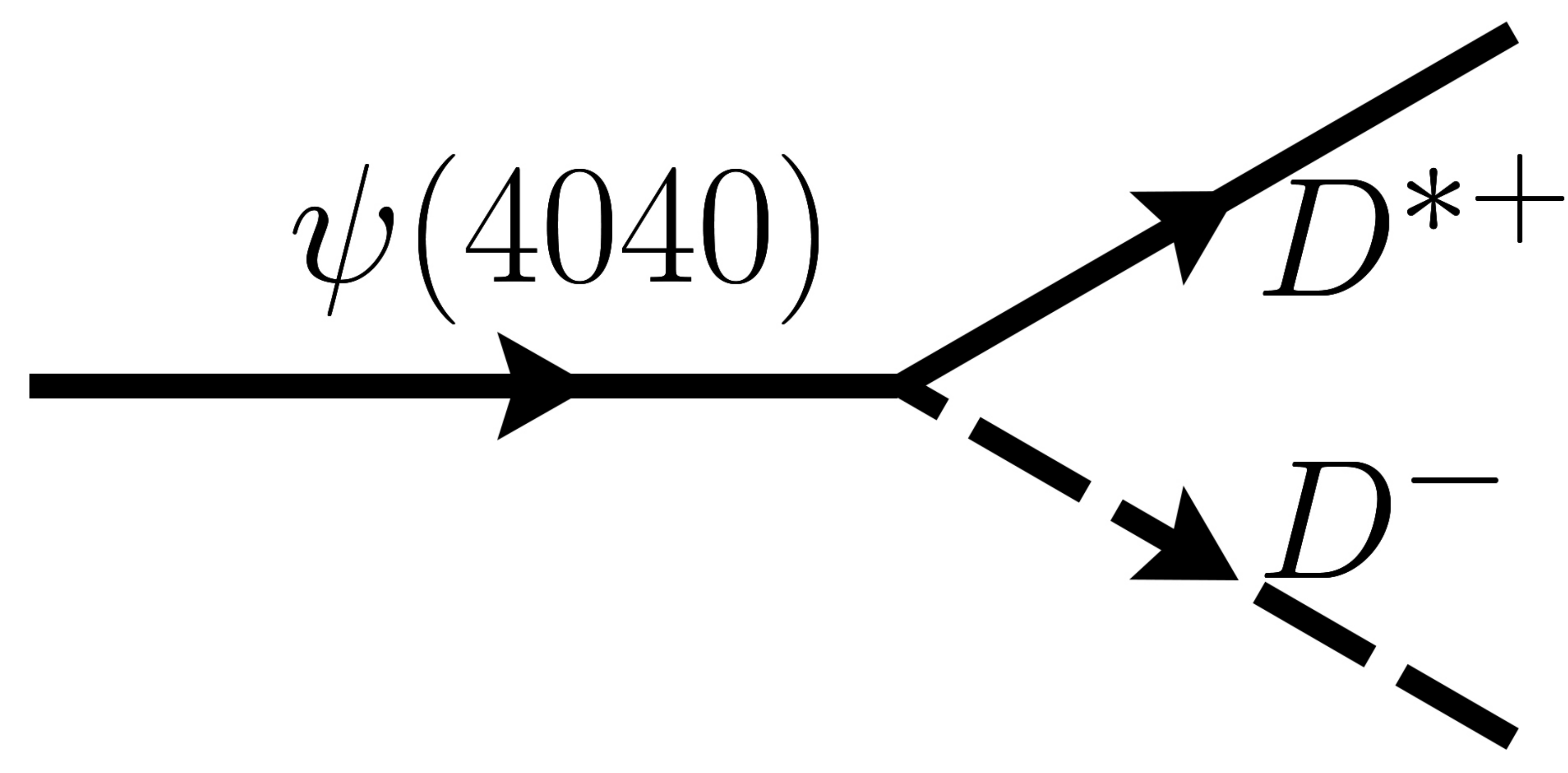}\\ 
\raisebox{5.5ex}{}&\raisebox{5.5ex}{$D^* D^*$}&\raisebox{5.5ex}{$\frac{2}{3}\frac{|\vec{k}|^3g^2_{\psi D^*D^*}}{\pi m^2_{D^*_c}}[2+\frac{|\vec{k}|^2}{m^2_{D^*}}]F_{\Lambda}$}&\includegraphics[width=0.215 \textwidth]{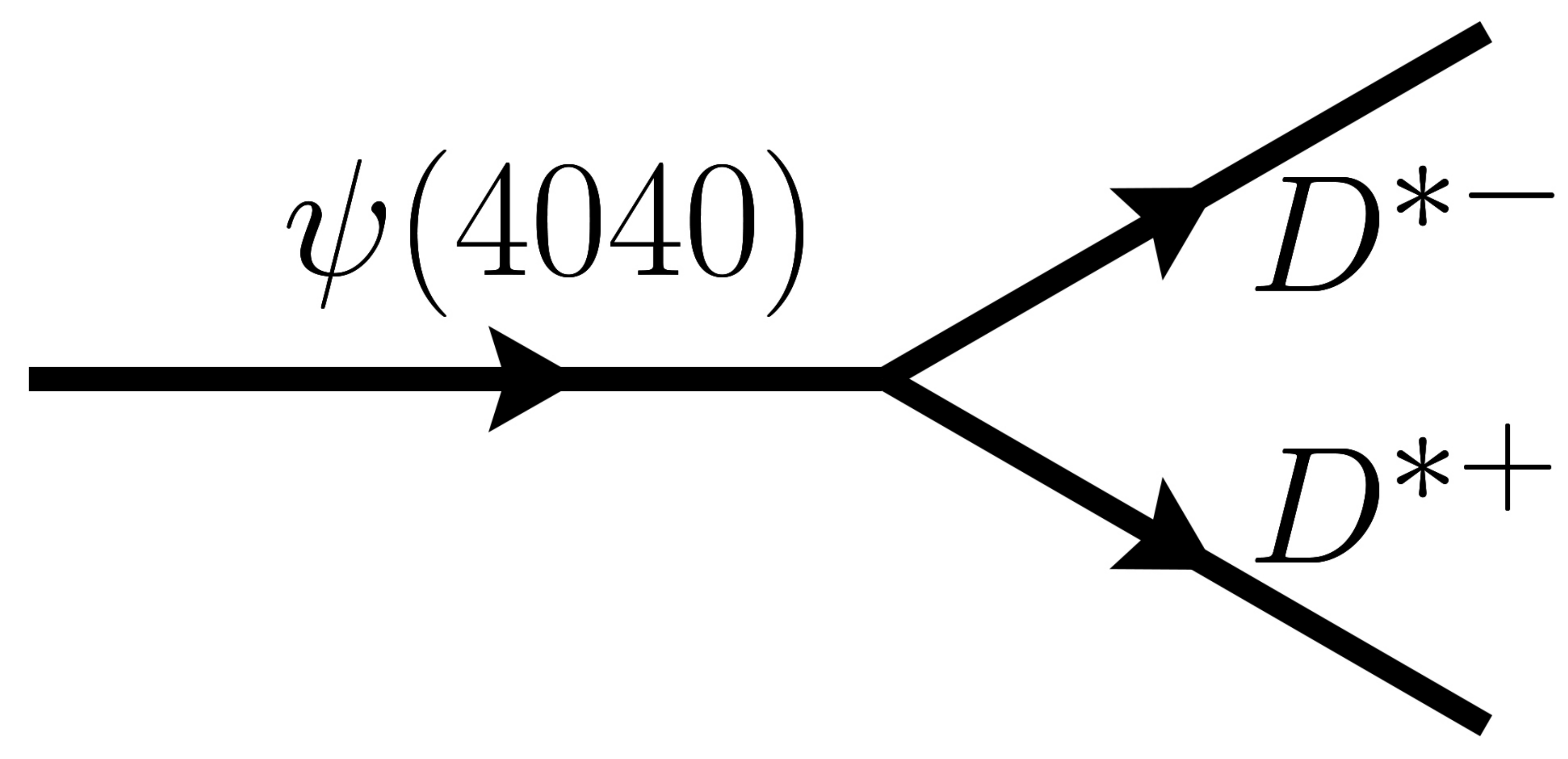}\\\hline 
\raisebox{5.5ex}{$\chi_{c1}(2P)$}&\raisebox{5.5ex}{$D D^*$}&\raisebox{5.5ex}{$\frac{2}{3}\frac{|k|g^2_{\chi_{c1}DD^8}}{8 \pi m^2}(3+\frac{|k|^2}{m^2_{D^*}})F_{\Lambda}$}&\includegraphics[width=0.215 \textwidth]{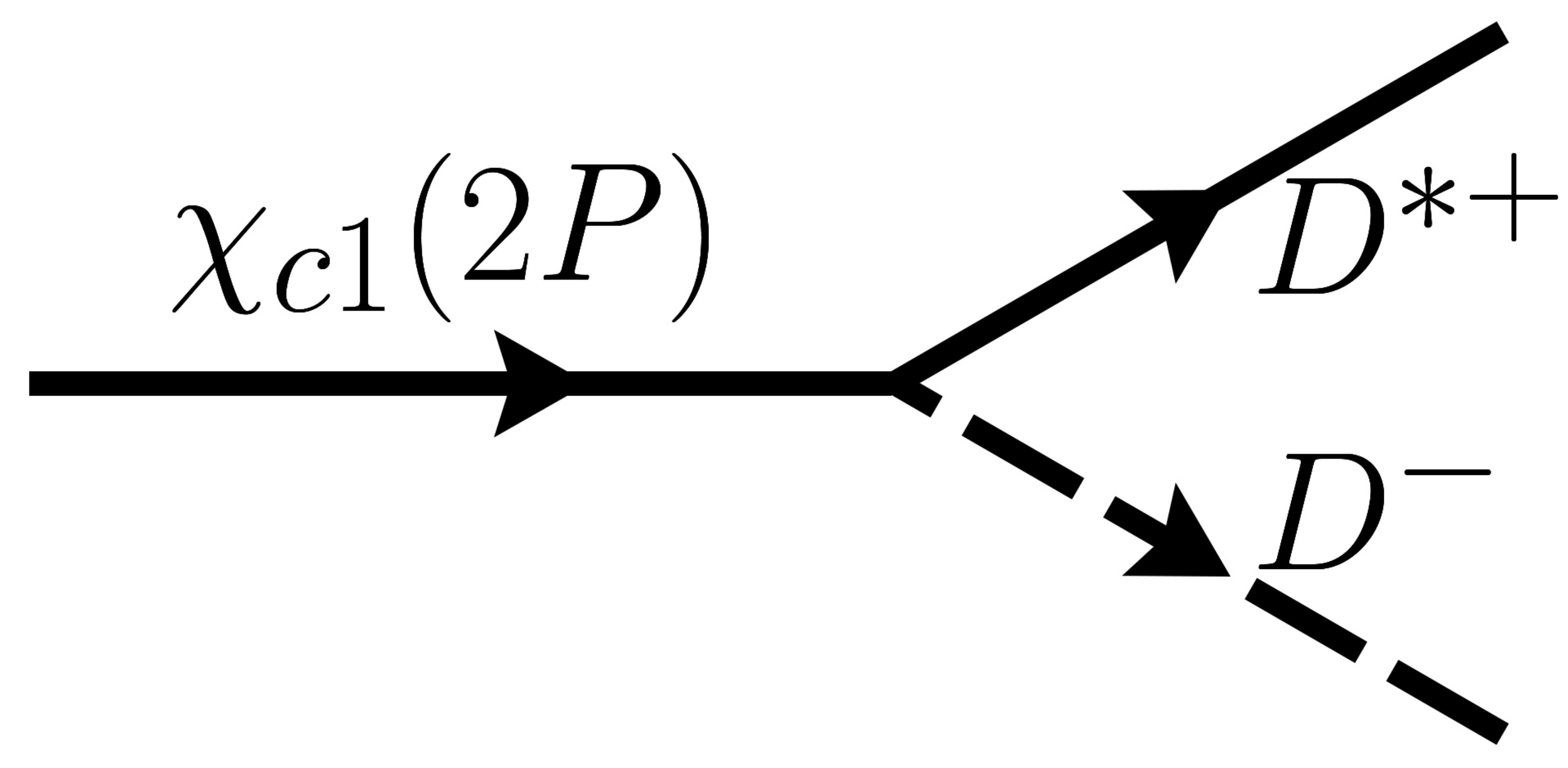}\\\hline \hline
\end{tabular}
}\caption{\label{tabwidth} Summary of the considered systems including information about main decay channels, theoretical expressions for the decay width, and some examples of the Feynmann diagrams.}%
\end{table}
The quantity $|\vec{k}|=\left(m^4+(m_A^2-m_B^2)^2-2(m_A^2+m_B^2)m^2\right)^{1/2}/2m$, with $m$ being the `running' mass of the resonance, is the three-momentum of one of the decay products, whose masses are $m_A$ and $m_B$. Moreover, our  model is regularized by the form factor $F_{\Lambda}$. Here we use the standard Gaussian function of the type
\begin{equation}
F_{\Lambda}(m)=e^{-2k^2(m)/ \Lambda^2}
\end{equation} 
with the $\Lambda$ parameter being an energy scale of the order of $1$ GeV.  

Next, we introduce the propagator of the standard $q\bar{q}$ seed state dressed by the mesonic quantum loops. The scalar part of it reads:
\begin{equation}
\Delta_f(m^2)=\left[m^2-M_{0,f}^2+ \Pi(m^2)+ i \varepsilon\right]^{-1} \hspace{0.2cm} ,
\end{equation}
where $f$ refers to the $K^*_0$, $\psi$ or $\chi_{c1}$ field. The quantity $M_{0,f}$, in the above, stands for the bare  mass of the corresponding seed state. Moreover, the function $\Pi(m^2)$ is the sum of all mesonic loop contributions. 

Finally, we are ready to define the spectral function, which is connected to the propagator by the following relation
\begin{equation}
d_{f}(m)=\frac{2m}{\pi}|Im \Delta_f(m^2)| \hspace{0.2cm}.
\end{equation}
The spectral function is nothing else than  the mass distribution of the unstable state. It determines the probability that the decaying resonance has the mass in range from $m$ to $m+dm$. Accordingly, it has to be normalized to unity:
\begin{equation}
\int \limits^{\infty}_{0}d_f(m)dm=1 \hspace{0.2cm}.
\end{equation}
Sometimes it is useful to calculate the partial spectral function which can be written as:
\begin{equation}
d_{f \rightarrow AB}(m)=\frac{2m}{\pi}|\Delta_f(m^2)|^2 m \Gamma_{f \rightarrow AB}(m) \hspace{0.2cm}.
\end{equation} 
This quntity can be understood as the probabilty that resonance $f$ has a mass between $m$ and $m+dm$ and decays into particular $f \rightarrow AB$ channel.
\section{Consequences of the model}
\begin{figure}[!h]
 \begin{center}
\includegraphics[width=0.9 \textwidth] {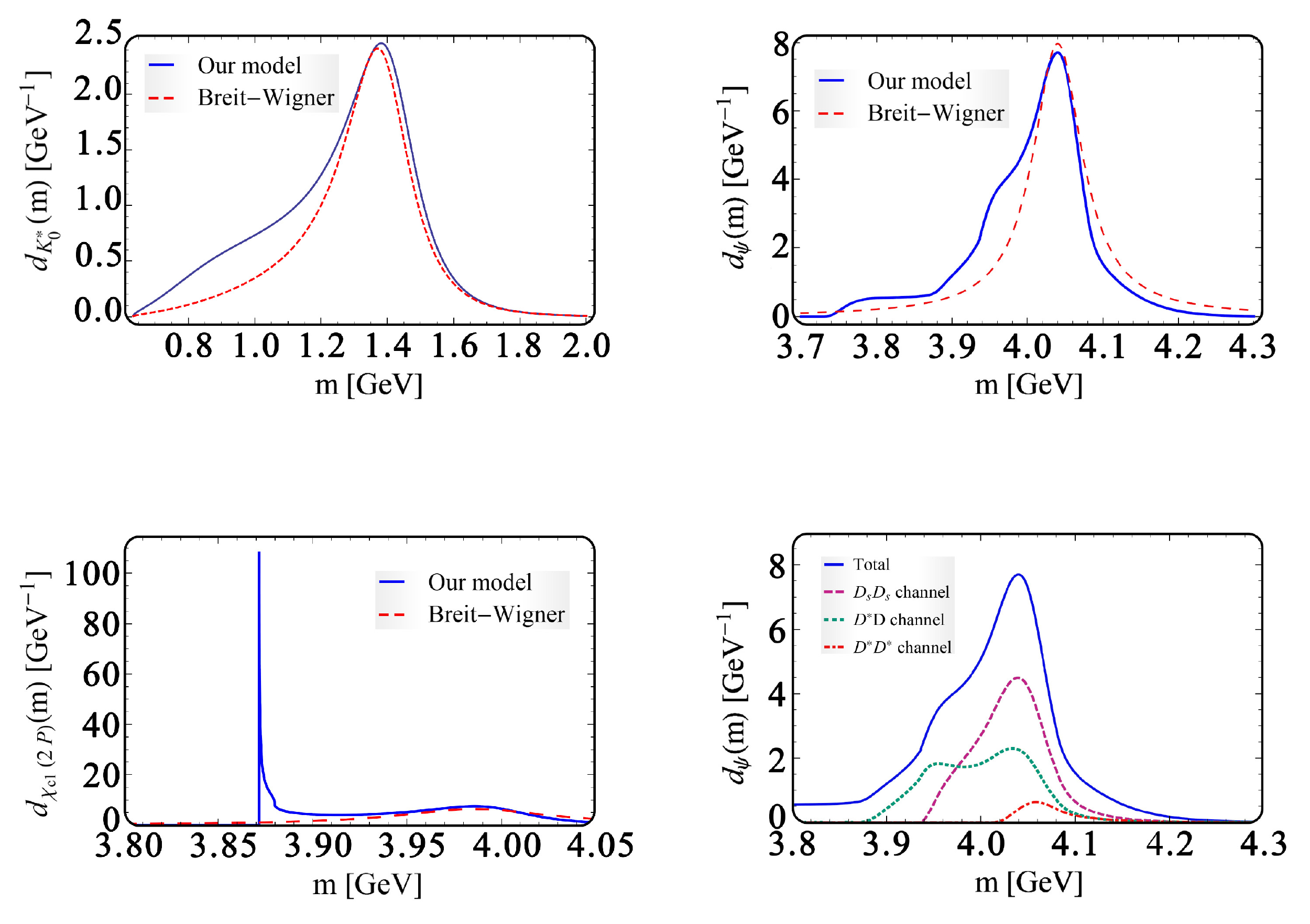}
\caption{\label{sfmp} Shape of the the total spectral function of the $K^*_0(1430)$ (left top), $\psi(4040)$ (right top) and $\chi_{c1}(2P)$ (left bottom) systems. The last panel (right bottom) presents the partial spectral function of $\psi(4040)$ resonance.  }
 \end{center}
\end{figure}

In Fig. \ref{sfmp} we present the spectral functions for the $K^*_0(1430)$, $\psi(4040)$ and $\chi_{c1}(2P)$ systems. For each case the obtained shape of the spectral function is compared with the standard Breit-Wigner distributions plotted for the parameters of the PDG \cite{pdg}. In addition, for the $\psi(4040)$ system a partial spectral functions are computed. 

Moreover, for each system we determined the positions of the poles. We found that in all cases two poles appear on the complex plane. Their coordinates are listed in Tab. \ref{polemp}.

\begin{table}[h]
\renewcommand{\arraystretch}{1.37}
\par
\makebox[\textwidth][c] { 
\par%
\begin{tabular}
[c]{c|c|cc|cc}\hline \hline
Seed&$q\bar{q}$ state &Pole for $q\bar{q}$ [GeV] &RS&Companion pole [GeV]&RS\\ \hline
$K^*_0$&$K^*_0(1430)$&$1.413-0.127 i$& II &$0.746-0.262i$&II \\ \hline
$\psi$&$\psi(4040)$&$4.053-0.039 i$&II& $3.934-0.030i$&II \\ \hline
$\chi_{c1}$&$\chi_{c1}(2P)$&$3.995-0.036i$&III& $3.87164-i\varepsilon$&II \\ \hline \hline
\end{tabular}
}\caption{\label{polemp} Poles for the system of $K^*_(1430)$, $\psi(4040)$ and $\chi_{c1}(2P)$ resonances.}%
\end{table}

For what concerns the $K^*_0(1430)$ resonance a deviation of the spectral function from the Breit-Wigner shape is visible in the low energy regime \cite{kappaMilena}. We found two poles on the complex plane: the expected one for the seed state, corresponding to the peak in the spectral function, and an  additional companion pole related to the enhancement. We assign this companion pole to $K^*_0(700)$ state, which very recently has been added to the summary table of the PDG (even if confirmation is still needed).  Within our approach we confirm its existence and explain its non-conventional nature: our interpretation of $K^*_0(700)$ as dynamically generated companion pole emerge naturally due to the mesonic $K \pi$ loops. The pole for $K^*_0(700)$ has been also determined in other works on the subject, see e.g. \cite{Rodas}. 

Similarly, the spectral function of the $\psi(4040)$ system is not covered by the Breit-Wigner shape. Again a dynamically generated enhancement appears in the lower energy regime. At first sight one may identify it with $Y(4008)$ state observed by the Belle Collaboration \cite{bell}. However, when comparing the coordinates of the additional pole a disagreement with the experimental data is visbile. From the plot of the partial spectral function one observes that this enhancement is mostly influenced by the $DD^*$ channel. A closer study reveals that the puzzling structure appears when consider the decay of $\psi(4040)$ into $J/ \psi \pi\pi$ channel through the intermediate $DD^*$ loop, see details in \cite{XMilena}.
Hence, $Y(4008)$ is not a real state but possibly only an effect of the strong coupling of $\psi(4040)$ to $DD^*$.  The existence of an additional pole is independent from this effect. 

Finally, we discuss the system with the $\chi_{c1}(2P)$ state. The shape of the spectral function is very peculiar.   The very high and extremely narrow peak at the lowest $D_0^*D_0$ threshold is the effect of dressing the $c\bar{c}$ seed state $\chi_{c1}(2P)$ by $D^*D$ loops. This peak is related to the $X(3872)$ resonance which in our approach emerges as a virtual companion pole, for details see Ref. \cite{psi4040Milena}. For the similar studies on the $X(3872)$, see e.g. Ref \cite{rupp}. 
\section{Conclusions}
Within our approach a mechanism of dynamical generation of companion poles has been used to explain the nature of some non-conventional mesons. We have shown that the light scalar state $K^*_0(700)$ can be interpreted as a companion pole of the heavier $K^*_0(1430)$ resonance. Similarly, the charmonium $X(3872)$ can be understood as virtual companion pole of the $\chi_{c1}(2P)$ state. For what concerns the system with the $\psi(4040)$ resonance, even if an additional companion pole exists, it can not be interpreted as $Y(4008)$. This is not a genuine resonance, but rather an enhancement apperaing when studying $\psi(4040)$ dressed by $DD^*$ loops.
\\

\textbf{Acknowledgements}: The author thanks F. Giacosa and P. Kovacs for the cooperations and useful discussions.

\end{document}